\documentclass[12pt]{article}
\usepackage{epsfig,graphicx}
\usepackage{fpcp03}

\def\beq{\begin{equation}}
\def\eeq#1{\label{#1}\end{equation}}
\def\eeqn{\end{equation}}

\def\beqa{\begin{eqnarray}}
\def\eeqa#1{\label{#1}\end{eqnarray}}
\def\eeqan{\end{eqnarray}}

\let\bar=\overbar

\def\half{\frac{1}{2}}

\def\Dslash{\not{\hbox{\kern-4pt $D$}}}
\def\dslash{\not{\hbox{\kern-2pt $\del$}}}

\def\msb{{\bar{\ssstyle M \kern -1pt S}}}

\def\s#1{\widetilde{#1}}

\def\BB0bar{B^0 {\overline B}^0}
\def\BB0dbar{B_d^0 {\overline B}_d^0}
\def\BB0sbar{B_s^0 {\overline B}_s^0}

\RequirePackage{xspace}

\usepackage{relsize}
\def\babar{\mbox{\slshape B\kern-0.1em{\smaller A}\kern-0.1em
    B\kern-0.1em{\smaller A\kern-0.2em R}}}

\def\s     {\ensuremath{s}\xspace}

\def\b     {\ensuremath{b}\xspace}

\def\Kbar  {\kern 0.2em\overline{\kern -0.2em K}{}\xspace}

\def\Kz    {\ensuremath{K^0}\xspace}
\def\Kzb   {\ensuremath{\Kbar^0}\xspace}
\def\KzKzb {\ensuremath{\Kz \kern -0.16em \Kzb}\xspace}
\def\Kp    {\ensuremath{K^+}\xspace}
\def\Km    {\ensuremath{K^-}\xspace}

\def\KpKm  {\ensuremath{\Kp \kern -0.16em \Km}\xspace}

\def\Dbar    {\kern 0.2em\overline{\kern -0.2em D}{}\xspace}

\def\Dz      {\ensuremath{D^0}\xspace}
\def\Dzb     {\ensuremath{\Dbar^0}\xspace}
\def\DzDzb   {\ensuremath{\Dz {\kern -0.16em \Dzb}}\xspace}
\def\Dp      {\ensuremath{D^+}\xspace}
\def\Dm      {\ensuremath{D^-}\xspace}

\def\DpDm    {\ensuremath{\Dp {\kern -0.16em \Dm}}\xspace}

\def\Bbar    {\kern 0.18em\overline{\kern -0.18em B}{}\xspace}

\def\BB      {\ensuremath{B\Bbar}\xspace} 
\def\Bz      {\ensuremath{B^0}\xspace}
\def\Bzb     {\ensuremath{\Bbar^0}\xspace}
\def\BzBzb   {\ensuremath{\Bz {\kern -0.16em \Bzb}}\xspace}
\def\Bu      {\ensuremath{B^+}\xspace}
\def\Bub     {\ensuremath{B^-}\xspace}

\def\BpBm    {\ensuremath{\Bu {\kern -0.16em \Bub}}\xspace}

\mathchardef\Upsilon="7107
\def\Y#1S{\ensuremath{\Upsilon{(#1S)}}\xspace}

\mathchardef\Deltares="7101
\mathchardef\Xi="7104
\mathchardef\Lambda="7103
\mathchardef\Sigma="7106
\mathchardef\Omega="710A

\def\Deltabar{\kern 0.25em\overline{\kern -0.25em \Deltares}{}\xspace}
\def\Lbar{\kern 0.2em\overline{\kern -0.2em\Lambda\kern 0.05em}\kern-0.05em{}\xspace}
\def\Sigbar{\kern 0.2em\overline{\kern -0.2em \Sigma}{}\xspace}
\def\Xibar{\kern 0.2em\overline{\kern -0.2em \Xi}{}\xspace}
\def\Obar{\kern 0.2em\overline{\kern -0.2em \Omega}{}\xspace}
\def\Nbar{\kern 0.2em\overline{\kern -0.2em N}{}\xspace}
\def\Xb{\kern 0.2em\overline{\kern -0.2em X}{}\xspace}

\newcommand{\tev}{\ensuremath{\mathrm{\,Te\kern -0.1em V}}\xspace}
\newcommand{\gev}{\ensuremath{\mathrm{\,Ge\kern -0.1em V}}\xspace}
\newcommand{\mev}{\ensuremath{\mathrm{\,Me\kern -0.1em V}}\xspace}
\newcommand{\kev}{\ensuremath{\mathrm{\,ke\kern -0.1em V}}\xspace}
\newcommand{\ev}{\ensuremath{\mathrm{\,e\kern -0.1em V}}\xspace}
\newcommand{\gevc}{\ensuremath{{\mathrm{\,Ge\kern -0.1em V\!/}c}}\xspace}
\newcommand{\mevc}{\ensuremath{{\mathrm{\,Me\kern -0.1em V\!/}c}}\xspace}
\newcommand{\gevcc}{\ensuremath{{\mathrm{\,Ge\kern -0.1em V\!/}c^2}}\xspace}
\newcommand{\mevcc}{\ensuremath{{\mathrm{\,Me\kern -0.1em V\!/}c^2}}\xspace}

\def\mus  {\ensuremath{\rm \,\mus}\xspace}

\def\mus        {\ensuremath{\,\mu{\rm s}}\xspace}    

\def\to                 {\ensuremath{\rightarrow}\xspace}

\def\pep2{PEP-II}

\def\gsim{{~\raise.15em\hbox{$>$}\kern-.85em
          \lower.35em\hbox{$\sim$}~}\xspace}
\def\lsim{{~\raise.15em\hbox{$<$}\kern-.85em
          \lower.35em\hbox{$\sim$}~}\xspace}

\xspace

\newcommand{\app}       [1]  {{Acta Phys.\ Polon.\ {\bf #1}}}

\newcommand{\arnps}     [1]  {{Ann.\ Rev.\ Nucl.\ Part.\ Sci.\ {\bf #1}}}

\newcommand{\epjc}      [1]  {\epjBase\ C~{\bf #1}}

\newcommand{\ijmpa}     [1]  {{Int.\ J.\ Mod.\ Phys.\ {\bf A{\bf #1}}}}

\newcommand{\nima}      [1]  {\nimBaseC~A~{\bf #1}}

\newcommand{\npb}       [1]  {\npBase\ B~{\bf #1}}
\newcommand{\npps}      [1]  {{Nucl.\ Phys.\ Proc.\ Suppl.\ {\bf #1}}}

\newcommand{\plb}       [1]  {\jplBase\ B~{\bf #1}}

\newcommand{\rpp}       [1]  {{Rep.\ Prog.\ Phys.\ {\bf #1}}}

\newcommand{\zpc}       [1]  {\zpBase\ C~{\bf #1}}

\def\jetset74   {\mbox{\tt Jetset \hspace{-0.5em}7.\hspace{-0.2em}4}\xspace}

\begin{document}
\begin{flushright}
EFI 03-31 \\
hep-ph/0306308 \\
\end{flushright}

\vskip 1cm

\Title{Phenomenological methods for unitarity triangle angles\footnote
{Invited talk at FPCP 2003, the Second Conference on Flavor 
Physics and CP Violation, Paris, France, June 3$-$6, 2003, to be published 
in the proceedings.}}
\bigskip


%
\label{GronauStart}

%
\author{ Michael Gronau\footnote{Permanent address:
Physics Department, Technion, Haifa, Israel.}\index{Gronau, M.} }

%
\address{Enrico Fermi Institute and Department of Physics\\
University of Chicago, Chicago, Illinois 60637
}

\makeauthor\abstracts{
I review several methods for determining the CKM phases 
$\alpha \equiv \phi_2$ and $\gamma \equiv \phi_3$ through CP asymmetry 
measurements in hadronic $B$ decays. The current status of these 
measurements and the near future feasibility of the methods are discussed.
}

\def \app{D_{\pi \pi}}
\def \b{{\cal B}}
\def \bbpp{\overline{{\cal B}}_{\pi \pi}}
\def \bea{\begin{eqnarray}}
\def \beq{\begin{equation}}
\def \bg{\bar \Gamma}
\def \bl{\bar \lambda}
\def \bo{B^0}
\def \ko{K^0}
\def \lesssim{\stackrel{<}{\sim}}
\def \largesim{\stackrel{>}{\sim}}
\def \bpb{\stackrel{(-)}{B^0}}
\def \cn{Collaboration}
\def \cpp{C_{\pi \pi}}
\def \eea{\end{eqnarray}}
\def \eeq{\end{equation}}
\def \ite{{\it et al.}}
\def \kpb{\stackrel{(-)}{K^0}}
\def \lpp{\lambda_{\pi \pi}}
\def \ob{\overline{B}^0}
\def \ok{\overline{K}^0}
\def \rpp{R_{\pi \pi}}
\def \rt{r_{\tau}}
\def \s{\sqrt{2}}
\def \half{\frac{1}{2}}
\def \3half{\frac{3}{2}}
\def \spp{S_{\pi \pi}}

\section{Introduction}
The Kobayashi-Maskawa model for CP violation \cite{KM} passed in a 
remarkable way its first crucial test in $B$ decays \cite{Sanda} 
when a large CP asymmetry was measured \cite{psiKs,Ford} in $B^0 
\to J/\psi K_S$ in agreement with expectations. The  
virtue of this gold-plated decay mode is the absence of hadronic 
uncertainties~\cite{MG,LP} in predicting the {\em mixing induced 
asymmetry} in terms of a fundamental phase parameter $\beta\equiv 
\phi_1$ of the Standard Model. This opens a new era, in which other 
CP asymmetries in $B$ and $B_s$ decays ought to be measured in 
order to test the KM mechanism in an unambiguous way. One would hope 
that this will lead to a point where deviations 
from the simple KM framework will be observed. 

Very optimistically, one is looking forward to signals of new physics 
\cite{GrWo,Hiller}, possibly exhibited by large deviations from $\sin 
2\beta$ in asymmetries for charmless strangeness changing penguin 
dominated $B^0$ decays to CP final states, such as 
$\phi K_S, \eta'K_s$ or $K^+K^-K_S$ \cite{Ford}. Moderate deviations 
from $\sin 2\beta$ are allowed within the KM framework, caused by  
amplitudes carrying a different weak phase. These amplitudes involve 
hadronic uncertainties \cite{SanMin}. In order to 
make a clear case for physics beyond the Standard Model, these 
effects must be carefully bounded \cite{GLNQ,CGR} using flavor symmetry 
arguments and input from experimental data.

A systematic study of the CP violating phase structure of weak quark 
couplings requires a measurement of the phase $\gamma\equiv \phi_3$, 
associated with {\em CP violation in direct decays}. The present 
uncertainties in this phase and in the phase $\phi_2 \equiv \alpha
= \pi - \beta - \gamma$, combining $\gamma$ with the $B^0-\bar B^0$ 
mixing phase $\beta$, are about $40^\circ$ \cite{CKMfit}, 
\beq\label{gamma}
38^\circ \le \phi_3 \equiv \gamma \le 80^\circ~,~~~~
78^\circ \le \phi_2 \equiv \alpha \le 122^\circ~,
\eeq
while $\beta$ is already known to within $7^\circ$, $20^\circ \le \beta 
\le 27^\circ$. 

The purpose of this review is to survey progress made 
recently in several promising methods for measuring
the phases $\gamma$ and $\alpha$. A major part of our discussion will 
concern charmless $B$ decays, in which interference of tree and penguin 
amplitudes leads to direct CP asymmetries. 
In Section 2 we study $\alpha$ in the CP asymmetry 
of $B \to \pi^+\pi^-$. The cleanest method, based on isospin symmetry 
alone, will be extended to a scheme using broken flavor SU(3). Flavor 
symmetries are applied  in Section 3 to $B\to K\pi$ decays in order to 
learn $\gamma$. The decays $B^{\pm}\to \eta (\eta')\pi^{\pm}$ are shown 
in Section 4 to potentially offer large CP asymmetries, which are also 
related to $\gamma$. Decays into charmed final states, $B \to DK$,
which are free of penguin amplitudes and hadronic uncertainties, 
will be discussed in Section 5, and Section 6 concludes this survey.       

\section{$\alpha$ from $B\to \pi^+\pi^-$}
The amplitude for $B\to \pi^+\pi^-$ consists of two terms with 
different weak and strong phases, 
\beq\label{TP}
A(\bo \to \pi^+ \pi^-) =  |T| e^{i \gamma} + |P| e^{i \delta}~.
\eeq
The weak phase $\gamma$ changes sign under charge-conjugation.
We use a convention \cite{con} in which ``tree'' 
($T$) and ``penguin'' ($P$) amplitudes involve CKM factors 
$V^*_{ub}V_{us}$ and $V^*_{cb}V_{cs}$, respectively. The time-dependent 
decay rate for an initially $B^0$ state is given by \cite{MG} 
\beq\label{pi+pi-}
\Gamma(B^0(t)\to\pi^+\pi^-) \propto e^{-\Gamma t}
\left [1 + \cpp\cos\Delta(m t) - \spp\sin(\Delta m t)\right ]~,
\eeq
where
\bea\label{spp}
\spp & = & \frac{2 {\rm Im}(\lpp)}{1 + |\lpp| ^2} = 
\sqrt{1- C^2_{\pi\pi}}\sin 2\alpha_{\rm eff} \ne \sin 2\alpha~,
\\
\label{cpp}
- A_{\pi\pi} & \equiv & \cpp = \frac{1 - |\lpp|^2}{1 + |\lpp|^2} \ne 0~,
\\
\label{lpp}
\lpp & \equiv & e^{-2i \beta} \frac{A(\ob \to \pi^+ \pi^-)}
{A(B^0 \to \pi^+ \pi^-)}~.
\eea

One measures three observables, the two asymmetries $-A_{\pi\pi}\equiv \cpp$ 
and $\spp$ (which also determine $\alpha_{\rm eff}$) and the charged-averaged 
rate $\bar\Gamma \equiv \frac{1}{2}[\Gamma(B^0) \to \pi^+\pi^-)+(\Gamma(\ob)
\to \pi^+\pi^-)]$. Since these observables 
depend on four parameters, $|T|, |P|,~\delta$ and $\gamma$, their 
measurements are insufficient for determining $\gamma$ or $\alpha$. Another 
input can be obtained by using isospin or broken flavor SU(3) symmetry.

\subsection{The isospin triangles}
One forms the amplitude triangle,
\beq\label{iso}
\s A(B^+ \to \pi^+\pi^0) - A(B^0 \to \pi^+\pi^-) = \s A(B^0 \to \pi^0\pi^0)~, 
\eeq
and its charge-conjugate, by measuring also the rates for $B^+\to \pi^+\pi^0,
~B^0\to\pi^0\pi^0$ and the charge-conjugate processes. The two triangles for 
$B$ and $\bar B$, which have a common base in an appropriate phase 
convention, $A(B^+ \to \pi^+\pi^0) = 
A(B^- \to \pi^-\pi^0)$, do not overlap because of possible CP asymmetries in 
$B \to \pi^+\pi^-$ and $B \to \pi^0\pi^0$. The mismatch angle between the 
two triangles $2\theta \equiv 2(\alpha_{\rm eff}-\alpha) = 
{\rm Arg}[A(B^0\to \pi^+\pi^-)A^*(\ob\to \pi^+\pi^-)]$ determines 
$\alpha$ \cite{GL}. One may 
include in this method a very small effect of an electroweak penguin 
amplitude which is related by isospin to the tree amplitude \cite{GPY}. 

The current world-average branching ratios, averaged over $B$ and $\bar B$,
are in units of $10^{-6}$ \cite{CGR,Bona}:
\beq
\b(\pi^{\pm}\pi^0) = 5.3 \pm 0.8~,~~~~\b(\pi^+\pi^-) = 4.6 \pm 0.4~,~~~
\b(\pi^0\pi^0) < 3.6~(90\%~{\rm c. l.}).
\eeq 
The sides of the isospin triangles which are difficult to measure are 
clearly $B^0\to\pi^0\pi^0$ and $\ob \to \pi^0\pi^0$.
The upper bound on the combined charge-averaged branching ratio may be 
used to set an upper limit on $\theta$. Assuming that the maximum value 
for $|\theta|$ is obtained when the two isospin triangles are right  
triangles, one finds \cite{GQ} $|\sin\theta| \le 
\sqrt{\bar\Gamma(\pi^0\pi^0)/\bar\Gamma(\pi^{\pm}\pi^0)}$. A  somewhat 
stronger 
upper bound \cite{GLSS}, depending on all three branching ratios, is 
obtained by avoiding this assumption which may not apply to actual branching 
ratio measurements. The current upper limit (at 90\% confidence level)
on $\b(\pi^0\pi^0)$, implying $\theta < 50^\circ$, is not yet useful.

Whereas a nonzero value has not yet been measured for $\b(\pi^0\pi^0)$,
it is encouraging to note that the present data on $B\to\pi^+\pi^-$ and
$B^\pm \to \pi^\pm \pi^0$, showing $2\bar\Gamma(\pi^\pm\pi^0)/\bar\Gamma
(\pi^+\pi^-) > 1$, already imply a nonzero value for $\b(\pi^0\pi^0)$. 
It follows from the two isospin triangles for $B$ and $\bar B$ that 
\cite{GLSS}
\beq
\b(\pi^0\pi^0) \ge \left( \sqrt{\frac{\b(\pi^\pm\pi^0)}{r_\tau}} -
\sqrt{\frac{\b(\pi^+\pi^-)}{2}}\right )^2
 > 0.2\times 10^{-6}~(90\%~{\rm c. l.})~,
\eeq
where $r_\tau \equiv \tau(B^+)/\tau(B^0) = 1.076 \pm 0.013$ \cite{LEP}.
This lower bound is expected to increase when errors in the 
above central values of $\b(\pi^+\pi^-)$ and $\b(\pi^\pm\pi^0)$ are 
reduced, and when a nonzero value is measured for $\cpp$ 
\cite{GLSS}. A recent estimate \cite{CGR}, 
$\b(\pi^0\pi^0)= (0.4-1.6)\times 10^{-6}$, shows that a direct 
signal for this mode may soon be measured if the branching ratio is near 
the upper end of this range.  
  
\subsection{Applying broken flavor SU(3)}
The amplitudes $T$ and $P$ occuring in $B\to\pi^+\pi^-$ may be related 
by flavor SU(3) to corresponding amplitudes $T'$ and $P'$ describing 
$B \to K\pi$ decays \cite{Zep,SW,GHLR}. Flavor SU(3) is not as good a 
symmetry as isospin, and symmetry breaking effects must therefore be 
included. A favored approach to SU(3) breaking, which can be checked 
experimentally, is based on an assumption that tree and penguin amplitudes 
factorize, as argued within QCD \cite{BBNS}. One way of implementing
this idea \cite{GR1} is to obtain the ratio $|P/T| \sim 0.3$ by assuming that 
$T$ can be related to $b \to \pi\ell\nu$ \cite{LR}, while $P$  may be 
related to a $\Delta S=1$ penguin amplitude which dominates 
$B^+\to K^0\pi^+$. Here we will describe an alternative approach 
\cite{con,Charles} which relies only on the second assumption.  

Writing 
\beq\label{Kpifac}
-A(B^+ \to K^0 \pi^+) = |P'| e^{i\delta} = |P| e^{i\delta} 
\frac{f_K}{f_{\pi}\tan\theta_c}~,
\eeq
we neglect a very small term with weak phase $\gamma$, ${\cal 
O}(|V^*_{ub}V_{us}/V^*_{cb}V_{cs}|)\approx 0.02$, disregarding its possible 
but unlikely enhancement by rescattering effects. This assumption and 
the factorization hypothesis may be tested by comparing $B^+\to K^0\pi^+$ 
with future measurements of $B^+\to \ok K^+$ \cite{FKNP}.

\begin{figure}
\centerline{\includegraphics[height=6.0cm]{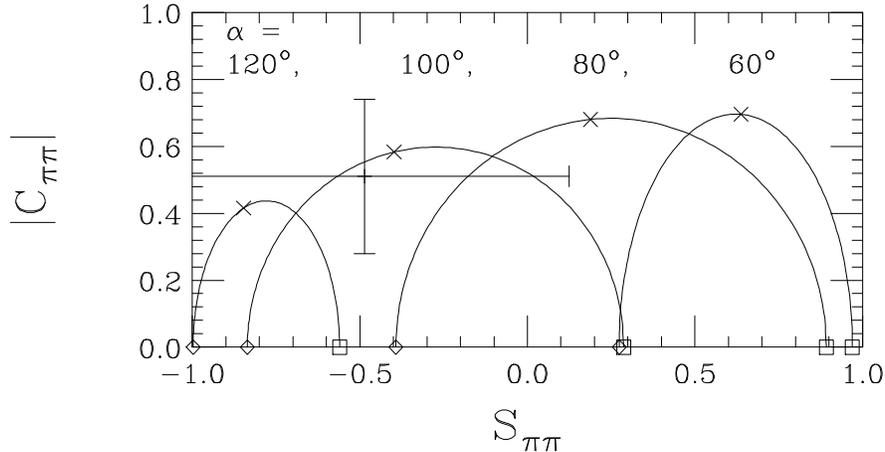}}
\caption{Plots of $|\cpp|$ versus $\spp$ for selected values
of $\alpha$, calculated for given values of $\beta$ and 
$\b(B^0 \to \pi^+\pi^-)/\b(B^+ \to K^0\pi^+)$. Small plotted points: 
$\delta =0$ (diamonds), $\delta = \pi$ (squares), $\delta = \pi/2$
(crosses).} 
\end{figure}  

Eqs.~({\ref{pi+pi-})--(\ref{lpp}) and Eq.~(\ref{Kpifac}) provide three 
measurables, $\spp,~\cpp$ and $\b(B^0 \to \pi^+\pi^-)/\b(B^+ \to K^0\pi^+)$,
which determine $|T/P|,~\delta$ and $\alpha$, assuming a given value for
$\beta$. Using $\beta = 24^\circ$ and $\b(B^0 \to \pi^+\pi^-)/\b(B^+ 
\to K^0\pi^+) = 0.23 \pm 0.03$ \cite{CGR,Bona}, we plot in Fig.~1 $|\cpp|$ 
versus $\spp$ for four selected values of $\alpha$ separated by $20^\circ$,
including the averaged measured values \cite{Sagawa}, $\spp = -0.49\pm 
0.61~(\chi=2.3),~\cpp = -0.51 \pm 0.23~(\chi=1.2)$, in which errors are 
inflated by rescaling factors as indicated. 
The plots are not very sensitive to the error in the above ratio of 
branching ratios. However, the present experimental error in $\spp$ is 
too large to constrain $\alpha$. 

\section{$\gamma$ from $B \to K\pi$}
The three decay modes $B^0\to K^+\pi^-,~B^+\to K^0\pi^+$ and $B^+\to 
K^+\pi^0$ are self-tagging, and can be used to learn $\gamma$ in 
more than one way. The three processes are dominated by penguin amplitudes, 
which can be related to each other by isospin alone. Subdominant electroweak 
penguin contributions are either color-suppressed or can be related 
to corresponding tree amplitudes by flavor SU(3), with small SU(3) breaking 
corrections. This provides two useful schemes, to be described here
\cite{otherSU3}, in which the 
number of measurables equals the number of unknowns, thus allowing a 
determination of the weak phase $\gamma$ between penguin and tree amplitudes.
The decay rate for $B^0\to K^0\pi^0$ is related to the above three processes
by an approximate sum rule \cite{Lipkin}. A current discrepancy in the 
sum rule, showing an enhancement in modes involving a neutral pion 
\cite{comb03}, indicates either a systematic underestimate of the efficiency 
for $\pi^0$ detection or new physics in $\Delta I=1$ transitions. 
The first effect may be canceled out by considering the product of two 
ratios of rates involving all four processes \cite{comb03}. We will  
mention briefly the information on $\gamma$ obtained when using this 
measurable.

\subsection{$B^0\to K^+\pi^-$ versus $B^+\to K^0\pi^+$}
Using isospin symmetry, we may write 
\beq
A(B^0 \to K^+ \pi^-) = |P'| e^{i\delta} - |T'| e^{i\gamma}~,
\eeq
where the first term describes also the amplitude (\ref{Kpifac}) for 
$B^+\to K^0\pi^+$. As mentioned, we neglect a very small color-suppressed 
electroweak penguin contribution. Denoting $r\equiv |T'|/|P'|$, we define
a charge-averaged ratio of rates
\beq\label{R0}
R_0 \equiv \frac{\bar \Gamma(K^\pm \pi^\mp)}{\bar \Gamma(K^0\pi^\pm)}
=  1 - 2r \cos\delta\cos\gamma +r^2 \ge \sin^2\gamma~,
\eeq
and a CP asymmetry
\beq\label{A0}
A_{\rm CP}(K^+\pi^-) \equiv \frac{\Gamma(K^-\pi^+) - \Gamma(K^+\pi^-)}
{\Gamma(K^-\pi^+) + \Gamma(K^+\pi^-)} = -2r\sin\delta\sin\gamma/R_0~,
\eeq
both of which are functions of $r,\delta$ and $\gamma$ \cite{GRKpi}.
The inequality in (\ref{R0}) holds for all values of $r$ and $\delta$, 
and would provide an interesting constraint on $\gamma$ if $R_0$ were 
smaller than one \cite{FM}. The present experimental value \cite{CGR}, 
$R_0 = 0.99 \pm 0.09$, is however consistent with 
one.

\begin{figure}
\centerline{\includegraphics[height=8.5cm]{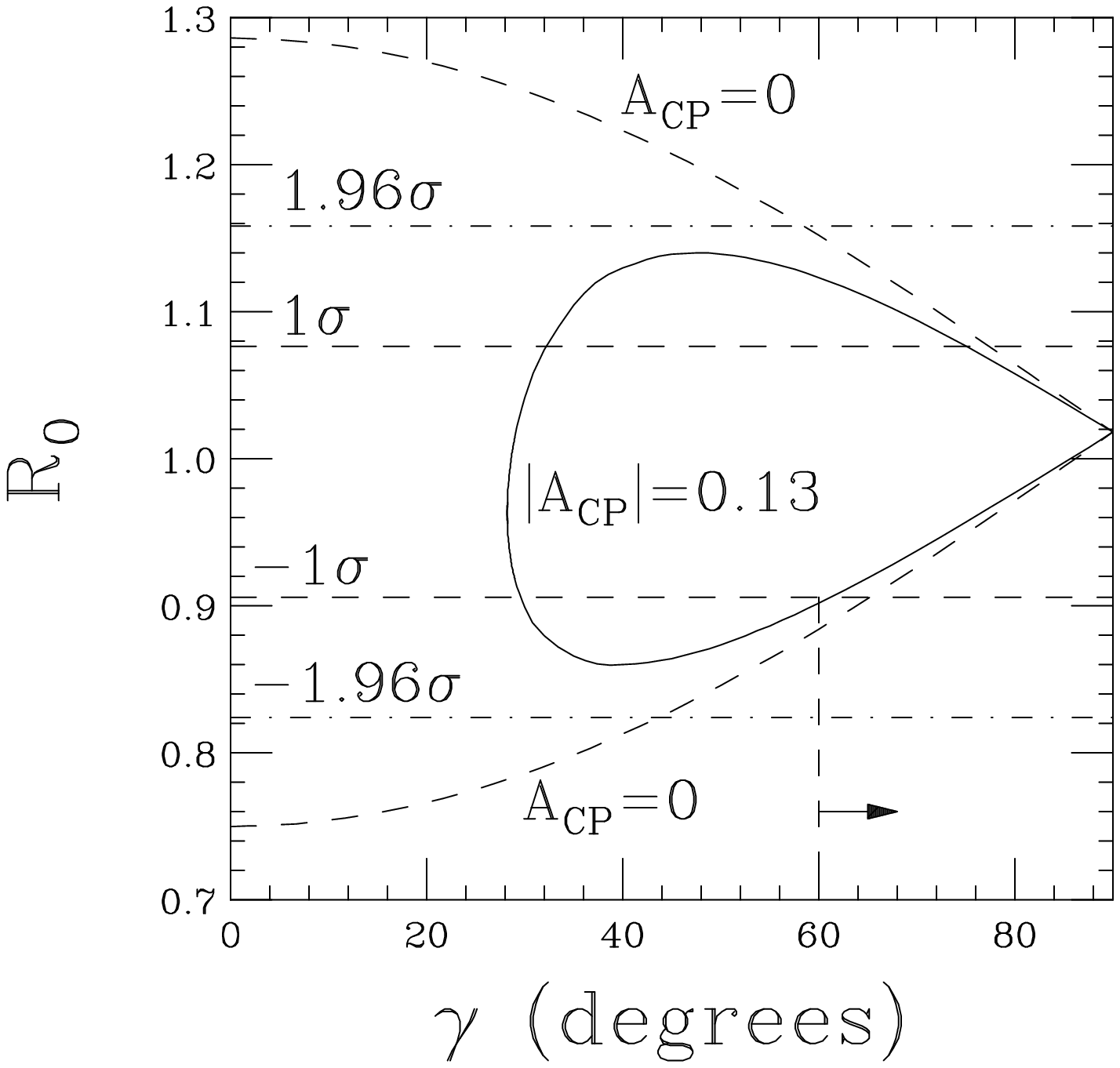}}
\caption{$R_0$ as a function of $\gamma$ for $r=0.13$ and 
$|A_{\rm CP}(K^+\pi^-)|=0.13$ (solid curve) or $A_{\rm CP}(K^+\pi^-)=0$ 
(dashed curve). Horizontal dashed lines denote 
$\pm 1\sigma$ experimental limits of $R_0$, while dot-dashed lines denote 
95\% c.l. limits. The lower branches of the curves correspond to $\cos\delta
\cos\gamma > 0$.}
%
\vskip 0.5cm
\centerline{\includegraphics[height=8.5cm]{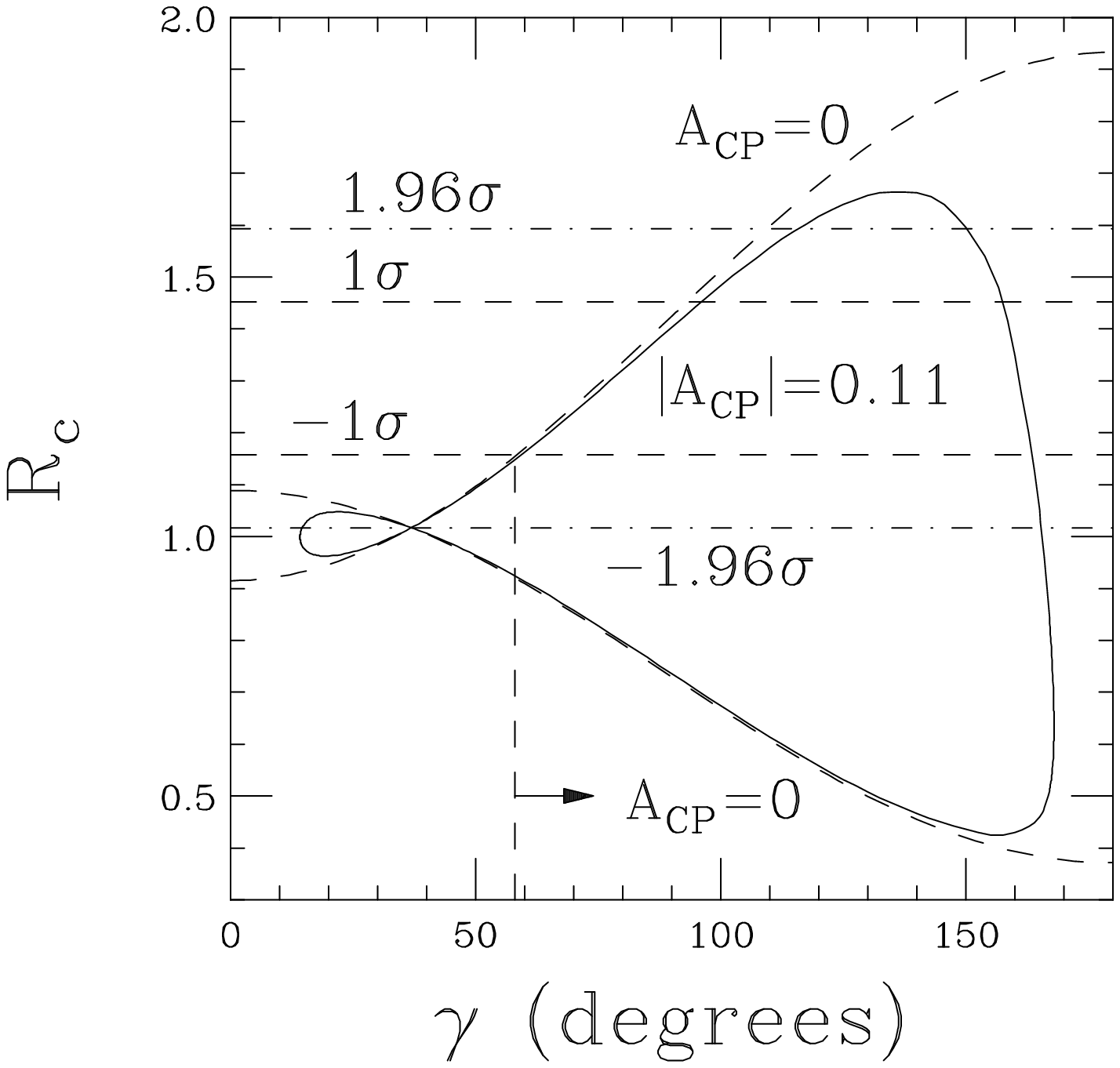}}
\caption{$R_c$ as a function of $\gamma$ for $r_c=0.22$ and 
$|A_{\rm CP}(K^+\pi^0)|=0.11$ (solid curve) or $A_{\rm CP}(K^+\pi^0)=0$ 
(dashed curve). Horizontal dashed lines denote 
$\pm 1\sigma$ experimental limits of $R_c$, while dot-dashed lines denote 
95\% c.l. limits. The lower branches of the curves correspond to 
$\cos\delta_c(\cos\gamma-\delta_{\rm EW})>0$.}
\end{figure}

We eliminate $\delta$ and plot in Fig.~2 \cite{comb03,JRLHC} $R_0$ versus 
$\gamma$ for the currently allowed $1\sigma$ range \cite{CGR,Bona}, 
$|A_{\rm CP}(K^+\pi^-)|$~$< 0.13$. 
The parameter $r$, obtained by comparing $B\to K\pi$ and $B \to \pi\pi$, 
involves a large theoretical uncertainty, $0.13 <r_{\rm th} < 0.21$ 
\cite{Kpi}. The most conservative constraints on $\gamma$ are obtained for 
$r=0.13$. We see that $\pm 1\sigma$ 
bounds on $R_0$ imply $\gamma > 60^\circ$, excluding about half of the
currently allowed values of $\gamma$ in Eq.~(\ref{gamma}). A slightly
more precise measurement of $R_0$ is needed in order to obtain new 
constraints on $\gamma$ at a $95\%$ confidence level.

\subsection{$B^+\to K^+\pi^0$ versus $B^+\to K^0\pi^+$}
The amplitude for $B^+$ decays into $|K\pi,I=3/2\rangle$ obtains
contributions from tree and electroweak penguin operators which are 
approximately proportional to each other \cite{NR}. The two physical 
amplitudes for charged $B$ decays may be written in terms of penguin ($P'$)
and tree ($T' + C'$) amplitudes, and the 
proportionality constant $\delta_{\rm EW}=0.65\pm 0.15$,  
\bea
\s A(B^+ \to K^+ \pi^0) & = & |P'| e^{i\delta_c} - 
|T'+C'|(e^{i\gamma}-\delta_{\rm EW})~,\\
-A(B^+ \to K^0 \pi^+) & = & |P'| e^{i\delta_c}~.
\eea
Denoting $r_c \equiv |T'+C'|/|P'|$, and defining a ratio of rates $R_c$
and an asymmetry $A_{\rm CP}$, one has
\bea\label{Rc} 
R_c & \equiv & \frac{2\bar \Gamma(K^\pm \pi^0)}{\bar \Gamma(K^0\pi^\pm)}
=  1 - 2r_c \cos\delta_c(\cos\gamma - \delta_{\rm EW}) + 
r_c^2(1 - 2\delta_{\rm EW}\cos\gamma + \delta^2_{\rm EW})~,
\\
A_{\rm CP}(K^+\pi^0) & \equiv & \frac{\Gamma(K^-\pi^0) - \Gamma(K^+\pi^0)}
{\Gamma(K^-\pi^0) + \Gamma(K^+\pi^0)} = -2r_c\sin\delta_c\sin\gamma/R_c~.
\eea

Using the ratio of branching ratios for $B^+\to \pi^+\pi^0$ and 
$B^+\to K^0\pi^+$, a range of values, $0.18 < (r_c)_{\rm th} < 0.22$, is 
obtained for the parameter $r_c$ \cite{GRL}. 
One then eliminates $\delta_c$ and plots in Fig.~3 \cite{comb03,JRLHC} 
$R_c$ versus $\gamma$ for the $1\sigma$ allowed range \cite{CGR,Bona},
$|A_{\rm CP}(K^+\pi^0)| < 0.11$. The values 
$r_c=0.22,~\delta_{\rm EW}=0.80$ are used for the most 
conservative bounds on $\gamma$. We see that the $1\sigma$ bounds 
\cite{CGR}, $R_c = 1.31 \pm 0.15$, already imply $\gamma > 58^\circ$. 
For $R_c>1$, the constraint on $\gamma$ is independent of the asymmetry 
\cite{NR}. Somewhat smaller errors in $R_c$ are needed for useful bounds 
on $\gamma$ at a $95\%$ confidence level.

\subsection{$B^0\to K^0\pi^0$}
One may also use the measured decay rate for $B^0/\bar B^0\to K^0\pi^0$.
An approximate sum rule between the four $B\to K\pi$ decay rates 
\cite{Lipkin} implies that, up to very small corrections, 
$R_c \approx R_n\equiv \bar\Gamma(K^\pm\pi^\mp)/2\bar\Gamma(K^0\pi^0)$. 
The current measurement \cite{CGR,Bona} $R_n = 0.81 \pm 0.10$ is almost 
$2\sigma$ {\em below} 1, whereas $R_c$ is $2\sigma$ {\em above} 1. 
This discrepancy may be caused either by new physics or by 
underestimating the $\pi^0$ detection 
efficiency. The latter effect may be canceled out by considering the quantity
$(R_cR_n)^{1/2}$, which is also described approximately by the 
right-hand-side of Eq.~(\ref{Rc}). 
The current $1\sigma$ bounds, $(R_cR_n)^{1/2} = 1.03 \pm 0.09$, imply  
$\gamma \le 77^\circ$ \cite{comb03}.

\section{The CP asymmetry in $B^+\to\eta \pi^+$}
$B^+$ decays into $\eta \pi^+$ and $\eta'\pi^+$ were anticipated to 
involve large CP asymmetries \cite{CPeta}, originating in an interference 
of tree and penguin amplitudes with comparable magnitudes.
Indeed, a recent BaBar result \cite{Baeta} favors a large asymmetry in
$B^+\to \eta \pi^+$. The $\eta$ and $\eta'$ correspond to octet-singlet 
mixtures
\beq
\eta  =   \eta_8 \cos \theta_0 + \eta_1 \sin \theta_0~,~~
\eta' = - \eta_8 \sin \theta_0 + \eta_1 \cos \theta_0~,
\eeq
with $\theta_0 = \sin^{-1}(1/3) = 19.5^\circ$.

\begin{figure}
\centerline{\includegraphics[height=8.0cm]{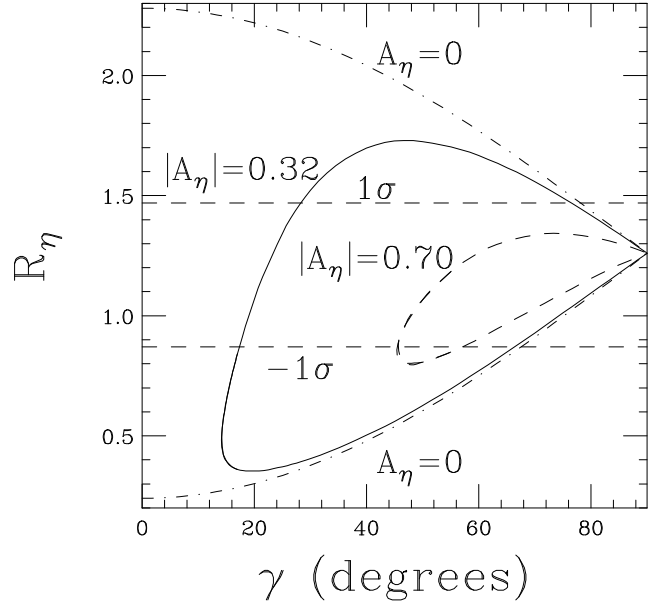}}
\caption{$R_\eta$ as a function of $\gamma$ for $r_\eta=0.51$ and $|A_\eta|=
0.70$ (solid curve) or $A_\eta=0$ (dashed curve). Horizontal dashed 
lines denote $\pm 1\sigma$ experimental limits of $R_\eta$.
The upper branches of the curves correspond to 
$\cos\delta_c\cos\gamma>0$}
\end{figure}
The amplitude for $B^+\to \eta\pi^+$, decomposed to its flavor 
contributions,
\beq
\sqrt{3}A(B^+\to \eta \pi^+) = |T+C|e^{i\gamma} + 
|2P + S|e^{i\delta_c}~,
\eeq
contains a tree amplitude which governs $B^+\to\pi^+\pi^0$,
\beq
\s A(B^+\to \pi^+\pi^0) = |T+C|e^{i\gamma}~.
\eeq
The amplitude of $B^+\to\eta\pi^+$ involves also the penguin amplitude $P$ 
occuring in $B\to \pi^+\pi^-$ multiplied by a factor two, and a small 
contribution from a new singlet term $S$. Denoting 
$r_\eta \equiv |2P +S|/|T+C|$ and neglecting the small $S$ term, the 
ratio of penguin to tree amplitudes in this process is given by \cite{CGR,
Ueta'}
\beq
r_\eta \equiv \frac{|2P +S|}{|T+C|} \largesim \frac{2|P|}{|T+C|} =
\frac{f_\pi\tan\theta_c}{f_K}\sqrt{\frac{2\b(K^0\pi^+)}{\b(\pi^+\pi^0)}}
= 0.51 \pm 0.04~.
\eeq
Defining a ratio of rates for the two processes, $R_\eta$, and an 
asymmetry in $B^\pm\to\eta \pi^\pm$, $A_\eta$, one has 
\bea
R_{\eta} &\equiv & \frac{3\bar\Gamma(\eta\pi^{\pm})}
{2\bar\Gamma(\pi^{\pm}\pi^0)} = 1 + r_\eta^2 + 
2r_\eta\cos\delta_c\cos\gamma~,\\
A_\eta & \equiv & \frac{\Gamma(\eta\pi^-) - \Gamma(\eta\pi^+)}
{\Gamma(\eta\pi^-) + \Gamma(\eta\pi^+)}
=-2r_\eta\sin\delta_c\sin\gamma/R_{\eta}~. 
\eea 

In Fig.~4 we eliminated $\delta_c$; we plot $R_\eta$ versus $\gamma$ 
for the measured $1\sigma$ range \cite{Baeta}, $A_\eta = -0.51 \pm 0.19$. 
The $1\sigma$ bounds \cite{CGR}, $R_{\eta} = 1.17 \pm 0.30$, do not 
constrain $\gamma$. We conclude that, although a large CP asymmetry 
measurement is very important, by itself it would not improve present 
constraints on $\gamma$. This would require a more precise measurement 
of $R_\eta$ and reducing the theoretical uncertainty in $r_\eta$. 

\section{$\gamma$ from $B^\pm \to D K^\pm$}
A theoretically clean method for determining $\gamma$, which avoids 
uncertainties in penguin amplitudes, was proposed some time ago \cite{GW}, 
using strangeness changing $B$ decays to neutral charmed mesons.  
One makes use of an interference between tree amplitudes in decays of 
the type $B^\pm\to DK^\pm$, from 
$\bar b \to \bar c u\bar s$ and $\bar b \to \bar u c\bar s$, for which the 
weak phase difference is $\gamma$. Several variants of this 
method were studied in the literature \cite{variants}. Here we 
will report the status of applying the original scheme \cite{BDK}, based 
on decays to $D^0$ flavor states and $D^0$ CP eigenstates, for which all 
the necessary observables were measured.

Denoting  by $r$ the ratio of amplitudes from $\bar b \to \bar u c\bar s$ 
and $\bar b \to \bar c u\bar s$, its weak phase by $\gamma$ and its 
strong phase by $\delta$, one finds for the two ratios of rates for 
even/odd CP and for flavor states, $R_\pm$, and for the two corresponding
CP asymmetries, $A_\pm$, expressions as in Section 3.1:
\bea
R_{\pm} & = & \frac{\Gamma(D^0_{{\rm CP}\pm} K^-) + 
\Gamma(D^0_{{\rm CP}\pm} K^+)}{\Gamma(D^0 K^-)} =
1 + r^2 \pm 2r\cos\delta\cos\gamma~,
\\
A_{\pm} & = & \frac{\Gamma(D^0_{{\rm CP}\pm} K^-) - 
\Gamma(D^0_{{\rm CP}\pm} K^+)}{\Gamma(D^0_{{\rm CP}\pm} K^-) 
+ \Gamma(D^0_{{\rm CP}\pm} K^+)} = 
\pm 2r \sin\delta \sin\gamma/R_{\pm}~.
\eea
In principle, the three independent observables determine $r,~\delta$
and $\gamma$. In practice, this may be difficult if $r$ is small.
Since this ratio involves the ratio of CKM factors $|V^*_{ub}V_{cs}
/V^*_{cb}V_{us}| = 0.4-0.5$ and a probably comparable color-suppression 
factor \cite{color}, a crude estimate is $r\sim 0.2$ \cite{BDK}. 

\begin{figure}
\centerline{\includegraphics[height=8.0cm]{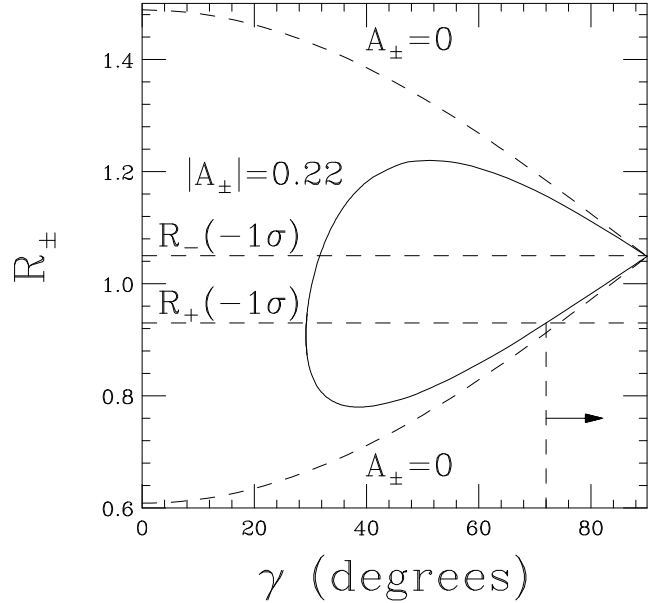}}
\caption{$R_{\pm}$ as functions of $\gamma$ for $r=0.22$ and 
$|A_{\pm}|=0.22$ (solid curve) or $A_{\pm}=0$ (dashed curve). Horizontal 
dashed lines denote $1\sigma$ experimental lower limits of $R_-$ and 
$R_+$. The lower and upper branches of the curves of $R_+$ 
and $R_-$, respectively, correspond to $\cos\delta\cos\gamma < 0$.}
\end{figure}

Averages of values measured by the Belle (Be) and BaBar (Ba) collaborations 
are \cite{dataDK} 
\bea
& & R_+ = 1.09 \pm 0.16~({\rm Be~\&~Ba}),~~R_- = 1.30 \pm 0.25~
({\rm Be})~\Rightarrow~r = 0.44^{+0.14}_{-0.22}~,\\
& & A_+ = 0.07 \pm 0.13~({\rm Be~\&~Ba}),~~~A_- = -0.19 \pm 0.18~
({\rm Be})~\Rightarrow~|A_{\pm}|_{\rm ave} = 0.11 \pm 0.11\,.
\eea
In order to obtain constraints on $\gamma$ we eliminate $\delta$, 
plotting in Fig.~5 $R_\pm$ versus $\gamma$ for allowed $A_\pm$. 
Using $1\sigma$ bounds on $R_\pm$ and $|A_{\pm}|_{\rm ave}$, we note that 
for these values of $R_-$, the ratios $R_+$ and $R_-$ are described 
by the lower and upper branches, respectively, corresponding to
$\cos\delta\cos\gamma < 0$. This implies a very strong 
constraint, $\gamma > 72^\circ$, and requires $\cos\delta < 0$ for
allowed values of $\gamma$. More precise measurements of 
$R_\pm$ are needed for constraints at a $95\%$ confidence level. 
One advantage of $B\to DK$ is that $R_+$ and $R_-$ are described by 
different branches of the curves. Since having both $R_\pm \ge 1$ is 
unlikely, either $R_+ < 1$ or $R_- < 1$ implies  
$\sin^2\gamma < 1$ which would provide also an upper bound on $\gamma$. 

\section{Conclusions and comparison with other approaches}
The phases $\alpha$ and $\gamma$ affect direct CP asymmetries, 
which require interference of two amplitudes with different weak 
and strong phases. Strong phases cannot be calculated reliably.
In $B\to \pi^+\pi^-, B\to K\pi$ and $B^\pm\to DK^\pm$ ratios of interfering 
amplitudes are typically $\sim 0.2-0.3$, whereas in $B^\pm\to \eta\pi^\pm$ 
and $B^\pm\to \eta'\pi^\pm$ ratios of penguin-to-tree amplitudes are larger, 
$\sim 0.5$ and $\sim 1$, respectively. These modes are therefore susceptible 
to large CP asymmetries. 

\begin{itemize}
\item The two $B\to \pi^+\pi^-$ asymmetry measurements
\cite{Sagawa} should converge before drawing any conclusion about $\alpha$. 
Indirect evidence for $B\to \pi^0\pi^0$ already exists from the two isospin 
related decays. A nonzero branching ratio for this decay mode is expected 
to be measured soon if the branching ratio is near the upper end of 
estimated values. This would provide a crucial step in performing a complete 
or a partial isospin analysis. 
\item Present experimental bounds on $B\to K\pi$ asymmetries are quite tight, 
and branching ratio measurements are sufficient for constraining $\gamma$ 
at a $1\sigma$ level to the range $60^\circ \le \gamma \le 77^\circ$. Some 
more statistics is needed for new bounds at a high degree of confidence level.
In particular, one awaits a resolution of a current 
discrepancy in a sum rule among the four $B\to K\pi$ rates, indicating an 
unexpected enhancement of the two processes involving a $\pi^0$.
\item Current measurements of $B^\pm \to D K^\pm$ constrain $\gamma$ at 
$1\sigma$ to $\gamma > 72^\circ$. Determining $\gamma$ in these processes 
(and in $B \to DK\pi$ \cite{BDK}) at a higher confidence level requires more 
accurate measurements of $R_\pm$, and would benefit from studying a variety 
of $D^0$ decay modes including $D^0\to K_S\pi^+\pi^-$ \cite{GGSZ}, in which 
CP and flavor quasi-two-body states interfere.
\end{itemize}

Finally, we wish to make several comments, comparing our approach to 
charmless $B$ decays with two other approaches. 
As we have stressed, our arguments were based primarily on flavor symmetries
and SU(3) breaking factors used to obtain ratios of amplitudes. The 
assumption that these effects are 
given to first order by factorization must be checked experimentally. 
In the absence of such tests, one chooses values for ratios of amplitudes 
that imply the most conservative constraints on $\gamma$. One must also
allow for small rescattering corrections which turn our curves into narrow 
bands, affecting constraints on $\gamma$ by a few degrees.
Future measurements will test both the effects of SU(3) breaking and those 
of rescattering, thus permitting less conservative and more restrictive
constraints. 

Two other approaches \cite{BBNS,KLS}, which we did not discuss,   
attempt to calculate within QCD magnitudes and strong phases 
of weak hadronic $B$ decay amplitudes. These calculations, which differ 
from one another in their predictions for strong phases, neglect 
incalculable higher order terms which may be large, and involve several 
phenomenological parameters depending on meson wave functions. As we 
have shown, knowledge of certain ratios of amplitudes suffices for 
constraining the weak phases. These ratios depend on SU(3) breaking 
factors, given in these approaches by the meson wave functions which must 
be determined from data. In this respect, our model-independent methods 
of learning weak phases are the essence of these more ambitious approaches.

\section*{Acknowledgments}
I am grateful to the Enrico Fermi Institute at the University of Chicago 
for its kind hospitality, and to C. W. Chiang, D. London, D. Pirjol, J. L. 
Rosner, N. Sinha, R. Sinha and D. Wyler for collaborations on the subjects 
mentioned here. I wish to thank J. L. Rosner in particular for providing 
Fig. 5, and for his useful comments on this manuscript. This work was 
supported in part by the United States Department of Energy through Grant 
No.\ DE FG02 90ER40560.

\def \ajp#1#2#3{Am.\ J. Phys.\ {\bf#1} (#3) #2}
\def \apny#1#2#3{Ann.\ Phys.\ (N.Y.) {\bf#1}, #2 (#3)}
\def \app#1#2#3{Acta Phys.\ Polonica {\bf#1}, #2 (#3)}
\def \arnps#1#2#3{Ann.\ Rev.\ Nucl.\ Part.\ Sci.\ {\bf#1}, #2 (#3)}
\def \art{and references therein}
\def \cmts#1#2#3{Comments on Nucl.\ Part.\ Phys.\ {\bf#1}, #2 (#3)}
\def \cn{Collaboration}
\def \econf#1#2#3{Electronic Conference Proceedings {\bf#1}, #2 (#3)}
\def \efi{Enrico Fermi Institute Report No.}
\def \epjc#1#2#3{Eur.\ Phys.\ J.\ C {\bf#1} (#3) #2}
\def \ib{{\it ibid.}~}
\def \ibj#1#2#3{~{\bf#1}, #2 (#3)}
\def \ijmpa#1#2#3{Int.\ J.\ Mod.\ Phys.\ A {\bf#1} (#3) #2}
\def \ite{{\it et al.}}
\def \jhep#1#2#3{JHEP {\bf#1} (#3) #2}
\def \jpb#1#2#3{J.\ Phys.\ B {\bf#1}, #2 (#3)}
\def \mpla#1#2#3{Mod.\ Phys.\ Lett.\ A {\bf#1} (#3) #2}
\def \nat#1#2#3{Nature {\bf#1}, #2 (#3)}
\def \nc#1#2#3{Nuovo Cim.\ {\bf#1}, #2 (#3)}
\def \nima#1#2#3{Nucl.\ Instr.\ Meth.\ A {\bf#1}, #2 (#3)}
\def \npb#1#2#3{Nucl.\ Phys.\ B~{\bf#1}  (#3) #2}
\def \npps#1#2#3{Nucl.\ Phys.\ Proc.\ Suppl.\ {\bf#1} (#3) #2}
\def \PDG{Particle Data Group, K. Hagiwara \ite, \prd{66}{010001}{2002}}
\def \pisma#1#2#3#4{Pis'ma Zh.\ Eksp.\ Teor.\ Fiz.\ {\bf#1}, #2 (#3) [JETP
Lett.\ {\bf#1}, #4 (#3)]}
\def \pl#1#2#3{Phys.\ Lett.\ {\bf#1}, #2 (#3)}
\def \pla#1#2#3{Phys.\ Lett.\ A {\bf#1}, #2 (#3)}
\def \plb#1#2#3{Phys.\ Lett.\ B {\bf#1} (#3) #2}
\def \prl#1#2#3{Phys.\ Rev.\ Lett.\ {\bf#1} (#3) #2}
\def \prd#1#2#3{Phys.\ Rev.\ D\ {\bf#1} (#3) #2}
\def \prp#1#2#3{Phys.\ Rep.\ {\bf#1} (#3) #2}
\def \ptp#1#2#3{Prog.\ Theor.\ Phys.\ {\bf#1} (#3) #2}
\def \rmp#1#2#3{Rev.\ Mod.\ Phys.\ {\bf#1} (#3) #2}
\def \rp#1{~~~~~\ldots\ldots{\rm rp~}{#1}~~~~~}
\def \yaf#1#2#3#4{Yad.\ Fiz.\ {\bf#1}, #2 (#3) [Sov.\ J.\ Nucl.\ Phys.\
{\bf #1}, #4 (#3)]}
\def \zhetf#1#2#3#4#5#6{Zh.\ Eksp.\ Teor.\ Fiz.\ {\bf #1}, #2 (#3) [Sov.\
Phys.\ - JETP {\bf #4}, #5 (#6)]}
\def \zpc#1#2#3{Zeit.\ Phys.\ C {\bf#1}, #2 (#3)}
\def \zpd#1#2#3{Zeit.\ Phys.\ D {\bf#1}, #2 (#3)}


\begin{thebibliography}{99}
\bibitem{Ueta'} The large ratio of penguin-to-tree amplitudes in $B^+\to\eta
\pi^+$ can be argued from U-spin \cite{CGR}. The corresponding ratio
in $B^+\to\eta' \pi^+$ is about one.
\end{thebibliography}
\end{document}